\documentclass[letter, conference]{IEEEtran}
\IEEEoverridecommandlockouts
\usepackage{cite}
\usepackage{amsmath,amssymb,amsfonts}
\usepackage{algorithm,algorithmic}
\usepackage{graphicx}
\usepackage{xcolor}
\usepackage{hyperref}
\usepackage{caption}
\usepackage{footnote}
\usepackage{threeparttable}
\usepackage[belowskip=2pt,aboveskip=3pt]{caption}
\usepackage{textcomp}
\usepackage{amsmath}
\usepackage{subfigure}
\usepackage{multicol,tabularx,capt-of}
\usepackage{hhline}
\usepackage{multirow}
\usepackage{listings}
\usepackage{booktabs}
\usepackage{etoolbox}
\usepackage{soul}
\usepackage[left=0.625in,right=0.625in,top=0.75in,bottom=1in]{geometry}

\appto\TPTnoteSettings{\scriptsize}

\def\BibTeX{{\rm B\kern-.05em{\sc i\kern-.025em b}\kern-.08em
    T\kern-.1667em\lower.7ex\hbox{E}\kern-.125emX}}
\begin{document}
\setlength{\textfloatsep}{0pt plus 2pt minus 2pt}

%\title{Mitigating 5G Registration Signaling Storm using Blockchain}
%\title{Achieving Security and Efficiency for NTN Handover in One Shot: A Group Handover Approach}
\title{Secure and Efficient Group Handover Protocol in 5G Non-Terrestrial Networks}

\makeatletter
\newcommand\notsotiny{\@setfontsize\notsotiny{7.7}{8.7}}

\newcommand{\linebreakand}{%
  \end{@IEEEauthorhalign}
  \hfill\mbox{}\par
  \mbox{}\hfill\begin{@IEEEauthorhalign}
}
\makeatother

\author{\IEEEauthorblockN{Bohan Zhang\IEEEauthorrefmark{1},\; Peng Hu\IEEEauthorrefmark{1}\IEEEauthorrefmark{2}, \;Ahmad Akbari Azirani\IEEEauthorrefmark{1}, \;Mohammad A. Salahuddin\IEEEauthorrefmark{1},\; Diogo Barradas\IEEEauthorrefmark{1},}

\IEEEauthorblockN{Noura Limam\IEEEauthorrefmark{1} and Raouf Boutaba\IEEEauthorrefmark{1}}

\IEEEauthorblockA{\IEEEauthorrefmark{1}David R. Cheriton School of Computer Science, University of Waterloo, Canada}
\IEEEauthorblockA{\IEEEauthorrefmark{2}National Research Council of Canada, Waterloo, Canada}
\{{b327zhan,\;peng.hu,\;a9akbariazirani,\;m2salahu,\;diogo.barradas,\;n2limam,\;rboutaba\}@uwaterloo.ca}}

\maketitle

  \begin{abstract}
The growing low-Earth orbit (LEO) satellite constellations have become an essential part of the fifth-generation (5G) non-terrestrial network (NTN) market. These satellites can enable direct-to-cell connectivity for mobile devices and support various applications with ubiquitous coverage for 5G and beyond networks. However, satellite-based NTNs bring several challenges to the 5G handover protocol design. The high mobility of satellites can lead to signaling storms and security compromises during handovers. This paper addresses these challenges by proposing a secure and efficient group handover protocol. The protocol's effectiveness is evaluated on a custom discrete-event simulator and compared against the baseline 5G handover scheme. The simulator is made publicly available.

\end{abstract}

  \section{Introduction}
\label{sec:introduction}

The wide-spreading low-Earth orbit (LEO) satellite constellations have become indispensable to the fifth-generation (5G) and beyond mobile networks for closing the connectivity gap. LEO-based non-terrestrial networks (NTNs) can enable the International Telecommunication Union  (ITU-R)'s envisioned framework for the new mobile communication generation, named IMT-2030, by supporting wide coverage, ubiquitous connectivity, and massive communication use case scenarios. {Integrating NTNs with 5G can extend the service coverage to a truly global scale and facilitate a wide range of applications, including disaster response and recovery, public safety, vehicular connectivity, and the Internet of Things (IoT).} In 5G and beyond NTNs, handover is pivotal in ensuring service continuity when user equipment (UE) signals must be transferred between orbiting non-terrestrial radio base stations, also known as gNBs.

The 3GPP standard, as specified in release 18 of TS 23.502, defines an Xn-based handover mechanism, where two gNBs can communicate, and an N2-based handover mechanism, where the 5G Access and Mobility Management Function (AMF) communicates with two gNBs. %These are expected to be used in NTN handover scenario, with satellite gNBs. 
However, satellite gNBs introduce various challenges, making these handover schemes, designed for terrestrial networks (TNs), difficult to apply in the integrated TN and NTN scenario \cite{3gpp33821}. First of all, due to the high dynamics and movements of the LEO satellites, numerous users will trigger handover requests around the same time, resulting in a large number of messages that need to be handled between UEs and satellites and between multiple satellites in a short period of time. This can easily cause a signaling storm\cite{gorbil2015modeling} in an NTN radio access network (RAN). This can be exacerbated by the expected support for massive machine-type communication (mMTC). Second, the signal strength used as an indicator for TN handover cannot be applied to the NTNs as the signal strength will experience a very small change within the satellite coverage\cite{3gpp33821}. Third, different from TNs, the ground-space and space-space links in NTNs are subject to much higher latency, which poses a significant limitation on the use of re-transmissions in an NTN handover scheme. Fourth, a satellite gNB has limited computing resources, posing real-world limitations on the processing capability of handover requests. Last but not least, an efficient NTN handover protocol must tackle all of the above challenges while operating securely, resisting attacks even in the case where UE devices may be compromised and weaponized. Therefore, it is critical to address these challenges when designing NTN handover protocols.

%Given the above challenges, the development of NTN handover solutions requires careful protocol design for ensuring efficiency and upholding security guarantees. 
Most existing works are based on TNs and address scenarios where handover poses a challenge, e.g., for UEs with high mobility that quickly change serving cells on platoons \cite{yan2022efficient} and high-speed rails \cite{yang2023fhap}. However, in a 5G NTN setting, the UE's mobility is overshadowed by the wide coverage of the satellite gNB~\cite{3gpp33821}. Moreover, due to unique challenges in 5G NTNs, such as high mobility of NTN gNBs relative to ground UEs, large number of UE connections, and security risks, traditional handover schemes for TNs cannot be applied to NTNs, and new handover protocols are needed. The existing NTN handover solutions are focused on the physical-layer handover \cite{juan2022handover}, handover procedure scheduling, and UE clustering \cite{user_cluster20, user_cluster22} without discussing the essential protocol design based on the latest 3GPP development. This paper bridges the gap in the NTN handover protocol design above the physical layer, while achieving efficiency and security. The main contributions are: 
%Peng's self-note: TO BE IMPROVED IN ANOTHER Iteration. Mention novelty and takeaways.
\begin{itemize}
%\item We propose a novel secure, efficient, and 3GPP-compliant Xn-based group handover protocol for NTNs.

\item We propose a {novel} secure, efficient, and 3GPP-compliant Xn-based group handover protocol for NTNs that mitigates signaling storms and malware security risks on UEs.

%\item Our proposed group handover protocol considers the characteristics of NTNs and mitigates the signaling storm issue and malware security risks on UEs. 

\item Our proposed protocol is implemented and validated through discrete-event simulations, where the key metrics such as success rate, message overheads, drop rate, and latency are significantly improved compared to the baseline 5G handover protocol \cite{peltonen2021comprehensive, 3gpp.23.502}.

\item We develop an open-source simulator where the NTN handover protocols can be implemented and evaluated.
%\textcolor{red}{[We should be explicit about the origin of the baseline protocol used for comparison]} 

%\item We validate through rigorous performance evaluation the efficiency of our protocol. The results can be reproduced. (Peng's NOTE: this is the common standard so we don't have to mention it as a contribution.)

%\item \textcolor{red}{[At least 2-3 distinct bullet points should be clearly listed here (and discussed clearly in the the experimental results discussion at least).]}
\end{itemize}

%\textcolor{red}{[In the contributions, we should also include the simulator that was developed and made publicly available for reproducibility]}
 
%The remainder of the paper is organized as follows. Section~\ref{sec:related-work} discusses the related work in 5G handover. The problem statement is in Section~\ref{sec:problems}. Our proposed group handover approach is described and analyzed in Section~\ref{sec:method}. Section~\ref{sec:implementation} discusses the simulation setup and evaluation results. The conclusive remarks and future work are presented in Section~\ref{sec:conclusions}.
  %\input{./content/3-background}
  \section{Related Work}
\label{sec:related-work}

The state-of-the-art handover solutions are mainly focused on TN scenarios, while NTN handover has recently emerged as a new research topic. In the following, we discuss the relevant works in TNs with UE mobility and recent advancements in NTN handover. Most of the following works are focused on the security challenges of group handover, while performance-related issues are less studied.  

In \cite{peltonen2021comprehensive}, a comprehensive security analysis demonstrates that 5G handover would not introduce confidential information leakage as long as the gNBs are not compromised. However, in NTN, this assumption is weakened and the mutual trust between satellites needs to be achieved in Xn-based handover. To accomplish this, a lightweight group key-based handover protocol is introduced in \cite{xue2020lightweight}, where satellites in a constellation negotiate a shared group key. The protocol allows a source satellite to send a handover ticket to a UE, and the UE can directly contact the target satellite using this ticket to achieve efficiency. However, the authors do not discuss how the target satellite can obtain necessary configuration messages from the source satellite, such as service session ID and C-RNTI.
%Their protocol removes the source satellite and target satellite communication and does not use asymmetric encryption, thus making the handover efficient. However, the protocol is vulnerable to a single point of failure because any satellite holding the group key can decrypt the handover ticket and obtain sensitive information from it. Also, the protocol allows the target satellite to obtain the previous session key and However, the protocol does not consider some necessary configuration messages required for handover, such as service session ID and C-RNTI. 
An identity-based encryption method is also proposed in \cite{kong2021achieving} that allows a UE to generate a symmetric key directly with a target gNB and ensure the source gNB does not know the future session key. 

In 5G networks, each AMF handles the tracking area where a UE is registered. Once UEs enter a new tracking area after handover, they need to be re-authenticated by AMF through the mobility registration update, as depicted in 3GPP TS 23.502 -4.2.2. In this case, the efficiency of processing many such re-authentication requests must be considered. Group authentication was proposed to be used before or after Xn-based handover \cite{lai2021novel, lai2022novel} and N2-based handover \cite{yan2022efficient, yang2023fhap} to reduce message overheads. In this case, UEs form groups and are assigned a group head (GH) to aggregate the information and perform group authentication. In NTNs, re-authentication is rarely expected after the handover as UEs do not move to another tracking area within a very short time. Also, the existing works assume that a GH is completely trusted and that group members do not leave a group during the handover procedure. These assumptions are unrealistic in the NTN handover scenarios due to the high mobility of LEO satellites and various malware attacks faced by UE devices especially IoT devices.

%However, these works \cite{li2020platoon, lai2021novel, lai2022novel} focus on authentication through an Xn-based handover, assuming that gNBs know the UE's Globally Unique Temporary UE Identity (GUTI). This assumption is not compliant with the 3GPP TS 33.501 in R18, which specifies that only AMF is known by a gNB. 

UE clustering algorithms for LEO satellite networks are discussed in \cite{user_cluster20, user_cluster22}, where the handover time scheduling problem is formulated to optimize performance metrics such as handover latency and quality of experience. However, the group handover protocols are not discussed in \cite{user_cluster20, user_cluster22} and a compromised GH can interrupt services of the group members. 
In our work, we consider the key characteristics of NTNs and propose an Xn-based handover protocol that enables NTN gNB to flexibly configure UEs to perform group/individual handover based on real-time situations without trusting a single GH. Our work will also address the signaling storm in handover procedures that can overload NTN gNB processors.

%Group handover (GHO) is a solution for 5G access networks that reduces HO signaling load. Satellite integrated 5G network studied by 3GPP \ref{3gpp33821}   is a good use case for the application of group handover. This is particularly relevant when dealing with scenarios where a large number of UEs are in use. That’s the reason why group handover in the LEO satellite network attracted the research community leading to the recent publication of works in this domain that we reference them and provide a brief discussion on their performance and security 

%\newpage
  %\;
%\newpage
\section{Problem}
\label{sec:problems}

\begin{figure}[t]
\centering
\includegraphics[width=0.75\columnwidth]{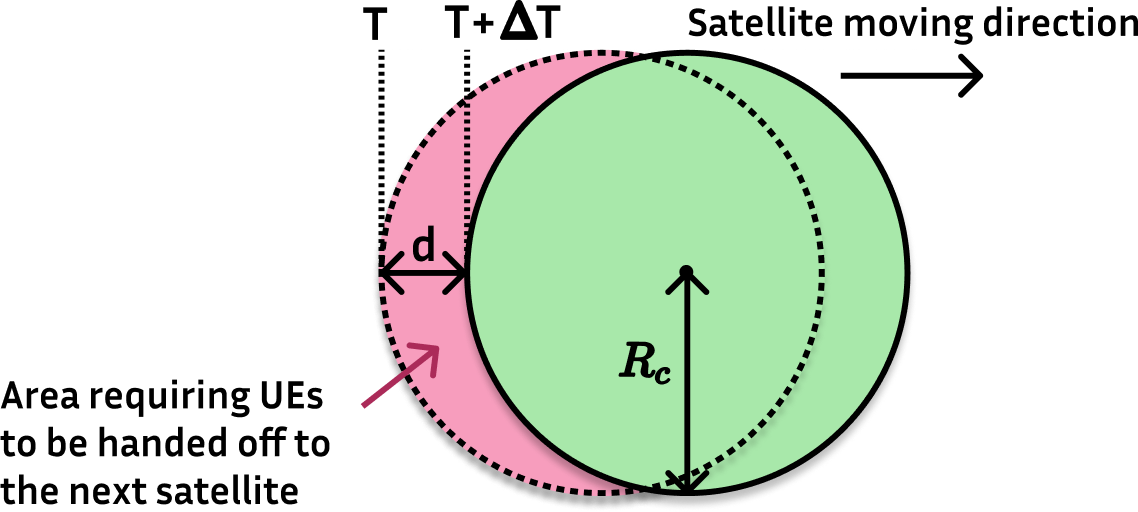}
\caption{Handover scenario with Earth moving cells}
\label{fig:3gppho}
\end{figure}
An important consideration based on the aforementioned challenges is the mobility of the LEO satellite. Compared to a 5G TN scenario where only a fraction of UEs requires handover within a short time, the rapid movement of LEO satellites (i.e., ${\sim}$7.56 km/s \cite{3gpp33821}) requires handover for a large number of UEs within a short period of time. Because LEO satellites cover much larger areas than ground stations, the excessive handover control signals from a large number of UEs can lead to a signaling storm, which can downgrade the network performance and delay the response to handover requests.
%\st{Furthermore, in 5G TNs, UEs have more time to stay in a cell intersection, giving the network more time to process handover messages. However, in 5G NTNs, the time available for a UE to handover is very short due to the continuous satellite movement.\mbox{\textcolor{red}{[redundant]}}} 
To avoid losing connectivity, UEs will usually retry their handover when they do not receive a response, which amplifies the signaling storm. To highlight this challenge, we calculate the expected number of UEs requiring handover within {a cell (i.e., cell-level handover depicted in 3GPP TS 38.300-9.2.3) in a given time} in the following paragraphs.

\vspace{1.5pt}\noindent\textbf{Handover scenario description:} Assume that the speed of a satellite is $V_{sat}$, cell radius is $R_{c}$, and all satellites have the same speed and cell radius. Variable $d$ represents the moving distance. Assuming the density of UEs is uniform, we calculate the number of UEs that must be handed off in a given period $\Delta T$. We denote {the intersection area of two circles} in Fig. \ref{fig:3gppho} as $A_{intersect}$ and the cell area as $ A_{circle} = \pi R_{c}^2$.
% $$ M = 2r^2 \arccos(\frac{d}{2r}) - d \sqrt{r^2-\frac{d^2}{4}} $$
% $$ d = v \Delta t $$
% $$ S = \pi r^2 - M$$
The red area requiring the satellite to hand off UEs, $A_{hand-off}=A_{circle} - A_{intersect}$, can be calculated as follows:

{\footnotesize
$$ A_{intersect} = 2R_{c}^2 \arccos({d}/{2R_{c}}) - d \sqrt{R_{c}^2-{d^2}/{4}} $$
$$A_{hand\text{-}off} = \pi R_{c}^2 - 2R_{c}^2 \arccos({d}/{2R_{c}}) + d \sqrt{R_{c}^2-{d^2}/{4}}.$$
}

Assuming a satellite connects to $N$ UEs uniformly {in a cell}, the UE density is ${N}/{A_{circle}}$. Within $\Delta T$, the number of UEs requiring to be handed off, $N_{hand\text{-}off}$, will be:  

{\footnotesize
$$N_{hand\text{-}off} = {N} \cdot A_{hand-off} / {A_{circle}} $$
%$$ = \frac{N}{\pi R_{sat}^2} ( \pi R_{sat}^2 - 2R_{sat}^2 \arccos(\frac{d}{2R_{sat}}) + d \sqrt{R_{sat}^2-\frac{d^2}{4}} )$$
$$ = N - {2N}\arccos({d}/{2R_{c}})/{\pi} + {dN}(\sqrt{4R_{c}^2 - d^2})/{2\pi R_{c}^2}.$$
}

In the Starlink setting\cite{starlink} where $R_{c}\approx12.07\; km$, $ V_{sat}\approx7.56 \;km/s$, and $N = 65519$ (the maximum C-RNTI value\cite{3gpp33821}), the UE density is about 36 UEs/km$^2$, requiring the satellite to handover $2.6 \times 10^4$ UEs/second. This challenge is exacerbated in the context of mMTC, where each LEO satellite serves a massive number of UEs (i.e., $10^6$ UEs/km$^2$, as depicted in 3GPP TS 22.261-6.4.2), requiring the satellite to handover $1.8\times 10^8$ UEs/second. Moreover, Sony indicated that handover signaling storms should be addressed in the first release of 3GPP NTN solution TS 38.821~\cite{sony}, while 3GPP indicates that this challenge will be addressed in the future~\cite{3gpp33821}.

  \section{Secure and Efficient Group Handover} 
\label{sec:method}

% \subsection{Cryptographic Primitives}

% \begin{enumerate}
%     \item EdDSA signature: Edwards-curve Digital Signature Algorithm uses a private-public key pair ($SK, PK = SK*G$) derived from an ellipic curve with generator $G$. To sign a message $msg$, the signer first computes $r =H(H(SK)||msg)$ and derives the signature $(R = r*G, S = r+H(R||PK||msg)*SK)$. \\
%     After the
% \end{enumerate}

\subsection{High-level Overview}
The objective of mitigating the handover signaling storm in an NTN is to reduce the workload of the source satellite (S-Satellite) by reducing its communication overhead. To achieve this goal, UEs near each other can form a group and  a GH is selected to perform handover from the source satellite on behalf of the group. However, a compromised GH can maliciously start an early, late, or PingPong~\cite{ghanem2012reducing} handover to disrupt the group member's service and increase the satellite's workload. To avoid this issue, we use  additive threshold secret sharing to allow the source satellite to know if the majority of the group members approve the GH's handover request. This scheme also ensures that the group handover succeeds even when some group members switch to inactive mode. 

In our solution, AMF is responsible for forming UE groups in its tracking area and delivering the group information to the UEs and satellites. Different secret shares are generated by the source satellite and delivered to each UE in the previous handover. The source satellite needs to select more than one GH from each UE group, called \textit{Group Aggregator} ($GA$), to decrease the risk of late handover and then deliver the hash commitment of the group member's share to each \textit{GA}. Furthermore, as the UE behavior is unpredictable, the source satellite must be capable of notifying or canceling the group-based handover based on real-time situations. For example, some UEs in a group switch to idle mode or de-register from the network, making the group unsuitable for group handover. 
During a group handover, a non-\textit{GA} member does not send the request to the satellite but publicly broadcasts its secret share to inform the \textit{GA} of its intention to handover. The \textit{GA} verifies the correctness of each received share by checking the hash commitment. After receiving enough shares above a given threshold, the \textit{GA} can generate the group handover ticket and initialize the group handover request to the source satellite. In such a way, the \textit{GA} cannot initialize an early or PingPong handover request. After receiving the group handover ticket, the source satellite first verifies the correctness of the group handover ticket. Then, it communicates with the target satellite (T-Satellite) to perform the handover for the group.

\subsection{Protocol Description}

\begin{figure}[t]
\includegraphics[width=0.95\columnwidth]{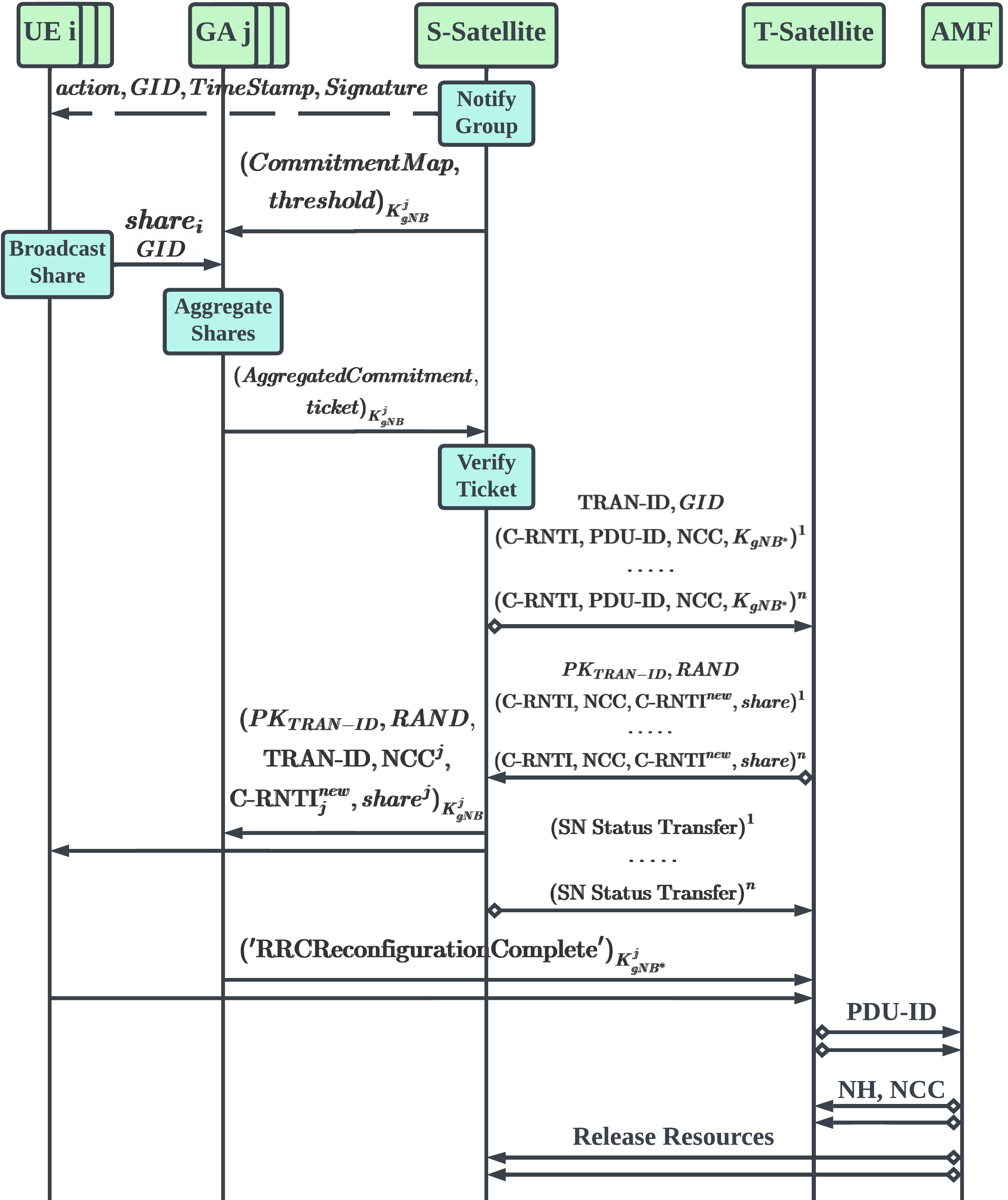}
\caption{Sequence diagram of the proposed NTN handover protocol (the symbol $\Diamond$ indicates a communication channel is secure; symmetric encryption with a key $K$ is denoted $(*)_K$)}
\vspace{-0.05cm}
\label{fig:groupworkflow}
\end{figure}

 \begin{algorithm}
 \caption{Source Satellite Processing Logic}\label{alg:two}
 \notsotiny
 \begin{algorithmic}[1]
 \renewcommand{\algorithmicrequire}{\textbf{Input:}}
 \renewcommand{\algorithmicensure}{\textbf{Output:}}
 \REQUIRE $RAN\text{-}ID$,  $RAND$, $GID$, $CommitmentShareMap$, $CommitmentMap$, $SK_{RAN\text{-}ID}$
% \\ \textit{Initialisation} :
  \STATE \textbf{Parallel Thread1:}
  \WHILE{Monitoring group $GID$ connected number}
  \STATE $nonce = RAN$-$ID||RAND||GID||TimeStamp$
  \IF{Group $GID$ is suitable for group handover}
  \STATE $action$ = `\texttt{SwitchToGroupHandover}'
  \STATE $m = action || nonce$
  \STATE $Sig$ = Sign($m$) using $SK_{RAN\text{-}ID}$
  \STATE Broadcast ($RAN$-$ID||GID||action||TimeStamp||Sig$)
  \STATE Decide $threshold$ based on the connected number
  \STATE Select Group Aggregators $AG_1,...AG_n$
  \STATE Send $(CommitmentMap, threshold)$ to each $AG_i$
  \ENDIF
  \IF{Group $GID$ does not need group handover}
  \STATE $action$ = `\texttt{CancelGroupHandover}'
  \STATE $m = action || nonce$
  \STATE $Sig$ = Sign($m$) using $SK_{RAN\text{-}ID}$
  \STATE Broadcast ($RAN$-$ID||GID||action||TimeStamp||Sig$)
  \ENDIF
  \ENDWHILE
  
  \STATE \textbf{Parallel Thread2:}
  \WHILE{Listening to group handover request}
  \STATE Receiving $(ticket, AggregatedCommitment)$
  \STATE Get share $share_1 ,..., share_t$ in $CommitmentShareMap$ based on $AggregatedCommitment$ and $CommitmentMap$
  \IF{$ticket$ == $share_1 \oplus$ ... $\oplus share_t$}
  \STATE Send group handover request to the target satellite
  \ENDIF
  \ENDWHILE
 \end{algorithmic} 
 \end{algorithm}

In this protocol, we assume satellites and AMF can securely communicate with each other. Considering a tracking area with a dense UE deployment, the corresponding AMF places UEs in different groups based on geometric location and UE attributes, such that group members will require a handover at a similar time\cite{user_cluster22}. AMF can decide to leave some UEs out of the group if their attributes are not suitable for group handover with neighboring UEs. If a UE is assigned to a group, it receives a groupID, $GID$, from the serving AMF during the registration or periodic update. AMF also informs the satellite of the group's geographic information and the expected number of UEs when the satellite performs the tracking area identity update. Each satellite with $RAN\text{-}ID$  generates a digital signature key pair $(SK_{RAN\text{-}ID}, PK_{RAN\text{-}ID})$. 

Fig. \ref{fig:groupworkflow} shows our group handover sequence diagram. The source satellite monitors the group's suitability for group handover. If suitable, the source satellite follows Alg.~\ref{alg:two}-Thread1  to notify the group members, selects the $GAs$, and sends each $GA$ the required threshold and the $CommitmentMap$, which contains the hash commitments of all group members. The group member should use $PK_{RAN\text{-}ID}$ to verify the notification message and then wait to perform the group handover. When a group member requests handover, it broadcasts the $GID$ and its secret share. The $GA$ follows Alg. \ref{alg:one} to verify the correctness of the share, aggregate the received shares, and send the handover request with an aggregated handover ticket to the source satellite. The source satellite follows Alg.~\ref{alg:two}-Thread2 to verify the correctness of the ticket and requests the target satellite to handover the group. The target satellite then returns the configuration information along with a random nonce $RAND_{TRAN\text{-}ID}$, its public key $PK_{TRAN\text{-}ID}$, and the new shares for group members. The target satellite saves the new shares and the corresponding commitments in a map called $CommitmentShareMap$. The source satellite then securely delivers the configuration, share, $RAND_{TRAN\text{-}ID}$ and $PK_{TRAN\text{-}ID}$ to group members individually. Each group member then attaches to the target satellite through random access and uses the new $K_{gNB^*}$ to achieve mutual trust.

\subsection{Protocol Analysis}
In this section, we analyze the proposed protocol's security and efficiency. We also discuss potential security and efficiency trade-offs of alternative solutions. 

\vspace{1.5pt}\noindent\textbf{Security:} Below, we analyze three important security aspects of our protocol.

\vspace{1.5pt}\noindent\textit{{Privacy preserving---}}The UE will only publicly broadcast the group ID and the secret. As there is no associated 5G identifier and the groupID cannot be used to distinguish group members, the UE will not leak private information that could be used, e.g., for a linkability attack. 

\vspace{1.5pt}\noindent\textit{{Avoid early/late/PingPong group handover---}}The GA cannot initialize group handover without most members agreeing to handover, so an early or malicious PingPong handover can be avoided. We assume at least one aggregator is benign, which means it will send the request as soon as the ticket is ready. Therefore, a malicious GA cannot postpone the handover.  

\vspace{1.5pt}\noindent\textit{{Replay resistant---}}A compromised party may try to forge/replay broadcasting information to maliciously inform UEs to perform group handover or cancel the group handover. The signature ensures that the broadcasting information cannot be forged in our protocol. The secret random nonce and timestamp ensure that the message cannot be replayed even if satellites periodically cover the UE.

 \begin{algorithm}
 \caption{Group Aggregator Processing Logic}\label{alg:one}
 \notsotiny
 \begin{algorithmic}[1]
 \renewcommand{\algorithmicrequire}{\textbf{Input:}}
 \renewcommand{\algorithmicensure}{\textbf{Output:}}
 \REQUIRE $threshold$, $CommitmentMap$, $GID$
% \\ \textit{Initialisation} :
  \STATE $ticket$ = 0
  \STATE $AggregatedCommitment$ = []
  \WHILE{Monitoring broadcasting information}
  \STATE Receive broadcasting message $(GID_i, Share_i)$
  \IF{$GID_i == GID$}
  \STATE $Commitment$ = Hash($GID||RAND||Share_i$)
  \IF{$Commitment$ exists in $CommitmentMap$}
  \STATE $ticket$ = $ticket \oplus Share_i$
  \STATE $i$ = $CommitmentMap.index(Commitment)$
  \STATE Add $i$ to $AggregatedCommitment$
  \ENDIF
  \ENDIF
  \STATE $Count$ = Length of $AggregatedCommitment$
  \IF{$Count > threshold$}
  \STATE Send handover request to the source satellite attaching ($ticket$, $AggregatedCommitment$)
  \ENDIF
  \ENDWHILE
 \end{algorithmic} 
 \end{algorithm}

\vspace{1.5pt}\noindent \textbf{Message overhead:} Consider a group $GID$ has $N_G$ UEs and $K_G$ GAs. To perform a group handover, the source satellite receives $K_G$ messages from GAs, one from the target satellite, and one from AMF, resulting in $2 +K_G$ messages. Compared to the 5G baseline handover, which requires the satellite to receive three messages per handover, our protocol significantly reduces the number of received messages from $3N_G$ to $2 + K_G$.

\vspace{1.5pt}\noindent \textbf{Security-efficiency trade-off:}
Below, we list some security concerns and alternative solutions that sacrifice efficiency.

\vspace{1.5pt}\noindent\textit{{Share distribution---}}In our protocol, the shares are generated by the target satellite and sent to UE by the source satellite. If the source satellite is compromised, the attacker who obtained shares can initialize an early handover to disrupt the service. To solve this, an alternative solution is that the source satellite distributes and sends the share and group handover notification to each UE during connection by replacing the broadcast notification message. However, this will add a computation overhead on the source satellite and decrease the overall performance, which was verified on our simulator.

\vspace{1.5pt}\noindent\textit{{Share broadcasting---}}In our protocol, the broadcast message from a UE is not protected and stays anonymous. A potential DoS attack on the aggregators may be carried out if the attacker knows the $GID$. The attacker can force the $GA$ to keep computing the hash. To mitigate this, one alternative solution is to enable a secure machine-to-machine communication channel between group members and GA. However, this will introduce identity leakage and lower efficiency due to decryption and channel establishment.

  \section{Performance Evaluation}\label{sec:implementation}
This section showcases the efficacy of our group handover protocol in mitigating signaling storms, and compares its performance to a baseline 5G handover protocol. %Before presenting our results, we introduce our custom simulator and describe the  experimental setup.

\subsection{Satellite Handover Simulator}
{We implement our protocol on a custom simulator that relies on the popular discrete-event simulation package \texttt{SimPy}, and provide a proof-of-concept evaluation.} Our simulator is designed for simulating handover protocols in NTNs, allowing us to perform experiments with an arbitrary number of satellites and UEs with mobility support. UEs are implemented as state machines where each state represents a different stage in the handover protocol. The simulator measures handover performance by tracking UE states at different timesteps and can generate animations. The simulator code is publicly available at: \texttt{\small \url{https://github.com/zbh888/LEOhandover}}.

\vspace{1.5pt}\noindent \textbf{Message processing model:} Fig. \ref{fig:queue} shows our message processing model. Our simulator computes the end-to-end communication delay by computing the following delays:
\begin{enumerate}
    \item \textit{{Propagation delay:}} This delay is computed based on the physical distance between the sender and receiver.
    \item \textit{{Transmission delay:}} The size of a control signal is less than 3,000 bytes, while 5G supports large bandwidth up to 20 Gbps. In these conditions, the delay is about 1 $\mu s$.
    
    \item \textit{{Queuing delay:}} Messages wait in a queue until a processor is available to handle handover requests, with the possibility of requests being dropped if the queue is full.%Each satellite has a message queue and multiple processors to process handover requests. Each processor takes a certain amount of time to process a certain message. The unprocessed messages are queued until a processor is available, and the satellite drops the handover request if the queue is full.
    
    \item \textit{{Processing delay:}} We divide message processing time into physical layer, logic execution, and cryptography tasks (encryption, decryption, signatures, and hashing), favouring batch hashing for efficiency purposes.  
    %We abstract the time for the processor to process a certain message to physical layer processing time, logic execution time, and cryptography-related processing time, including encryption, decryption, signatures, and hashing. We assume that batch hashing is more efficient than doing single message hashing.
\end{enumerate}

% To study the signaling storm problem during satellite handover, we implement our own simulator based on the popular discrete-event simulation (DES) package SimPy. By abstracting resources, our simulator supports simulating large-scale handover scenario to provide valuable insights to signaling storm problem. Our simulator is designed specifically to simulate the handover protocol in NTN. It is capable to simulate the deployment of extensive number of satellites and UE devices while also accommodating mobility support. The UEs are implemented as a state machine representing different state in handover protocol. By utilizing the UEs' states at different timestamp, the simulator can measure the handover performance and generate simulation animation to help understand the problem.

%DES is a method to simulate systems where events occur at specific, separable instances in time. It simulates the behaviour and performance of a system by computing system state after each event. 

% \begin{enumerate}
%     \item INACTIVE: The UE lost connection.
%     \item ACTIVE: Connecting to Satellite. Configured to perform non-group handover.
%     \item WAITING\_RRC\_RECONFIGURATION: The UE has sent a handover request and is waiting to receive RRC re-configuration.
%     \item RRC\_CONFIGURED: The UE received the RRC re-configuration
%     \item WAITING\_RRC\_RECONFIGURATION\_COMPLETE\_RESPONSE: The UE has sent an attachment request to the satellite and is waiting for the response.
% \end{enumerate}

\subsection{Experimental Setup}
\label{subsec:setup}

We now outline our scenario and introduce the configurations for the baseline handover and group handover protocols.
%We start by describing the scenario considered in our experiments, and introduce the configurations of the baseline handover and group handover protocols.

\begin{table}[!t]\centering
\caption{Simulation parameters and settings}
\label{table:setting}
\scriptsize
\begin{tabular}{lr|lr} 
  \toprule
  \textbf{Parameter} & \textbf{Value} & \textbf{Parameter} & \textbf{Value} \\ 
  \midrule
  {Satellite speed} & 7.56 km/s & {HO timeout} & 30 ms \\ 
  
  {Footprint radius} & 25 km &  {GHO timeout} & 35 ms  \\ 
  
  {Inter-satellite distance} & 30 km & {Square width} & 1 km  \\ 
  
  {Inter-satellite delay} & 1 ms & {Physical-layer processing time} & 0.05 ms  \\ 
  
  {Ground-satellite delay} & 3 ms & {Logic execution time} & 0.05 ms  \\ 
  
  {5G core-satellite delay} & 10 ms & {Encryption/Decryption} & 0.1 ms  \\ 
  
  {Transmission delay} & 1 $\mu s$ & {Signature sign/verify} & 0.3 ms \\
  
  {Message queue size} & 500 & {Single hash} & 0.05 ms \\
  
  {Max re-transmission} & 15 & {Batch hash} & 0.1 ms \\
  
  {Packet size} & 3,000 bytes & {Number of processors} & 4 \\
  \bottomrule
\end{tabular}
\end{table}

\begin{table*}[htb!]
    \begin{threeparttable}
  \centering
  \caption{Simulation results (baseline HO vs. GHO)}
  \scriptsize
  \label{table:res}
  \renewcommand{\arraystretch}{1.2}
  \begin{tabular}{c|rr|rr|rr|rr|rr|rr}
    \toprule
    \multirow{2}{1cm}{\textbf{UE Number}} & 
        \multicolumn{2}{c}{\textbf{Success Rate (\%)}} & 
        \multicolumn{2}{c}{\textbf{Total Messages}} &
        \multicolumn{2}{c}{\textbf{UE Messages}} &
        \multicolumn{2}{c}{\textbf{Drop rate (\%)}} &
        \multicolumn{2}{c}{\textbf{Successful UE WT (ms)\tnote{*}}} &
        \multicolumn{2}{c}{\textbf{Failed UE WT (ms)\tnote{**}}} \\

    % \hline
    % \textbf{Inactive Modes} & \textbf{Description}\\
    %\cline{2-13}
    & \textbf{HO} & \textbf{GHO} 
    & \textbf{HO} & \textbf{GHO} 
    & \textbf{HO} & \textbf{GHO} 
    & \textbf{HO} & \textbf{GHO} 
    & \textbf{HO} & \textbf{GHO} 
    & \textbf{HO} & \textbf{GHO} \\
    %\hhline{~--}
    \midrule
    % Success Rate (\%) | Total Messages | UE Messages |  Drop Rate (\%) | Success UE Waiting Time | Failure UE Waiting Time
    10,000 & 100.00 & 100.00    & 30,000 & 16,558  & 10,000 & 4,787  & 0.00 & 0.00        & 8.99±0.00 & 16.1±0.18 & --- & --- \\ \hline
    20,000 & 100.00 & 100.00    & 60,000 & 26,820  & 20,000 & 5,007  & 0.00 & 0.00          & 9.24±0.00 & 19.45±0.15 & --- & ---\\ \hline
    30,000 & 100.00 & 100.00    & 102,060 & 36,938 & 42,060 & 5,089  & 0.98 & 0.00      & 27.94±0.08 & 23.08±0.13 & --- & ---\\ \hline
    40,000 & 81.25  & 100.00    & 414,555 & 47,379 & 349,456 & 5,355 & 74.53 & 0.00      & 214.23±0.70 & 25.55±0.12 & 858.48±3.35 & --- \\ \hline
    50,000 & 66.56   & 100.00   & 602,981 & 67,275 & 536,317 & 12,497 & 82.28 & 0.00     & 246.20±0.71 & 71.25±0.21 & 887.84±5.22 & --- \\ \hline
    60,000 & 56.26  & 92.86     & 777,770 & 121,982 & 710,143 & 47,664 & 86.17 & 1.09 & 263.77±0.72 & 338.67±0.93 & 901.55±1.66 & 437.15±2.03\\ \hline
    70,000 & 48.70  & 82.70     & 943,105 & 137,979 & 874,797 & 58,135 & 88.57 & 2.60 & 274.47±0.72 & 482.52±1.19 & 909.38±1.40 & 568.81±1.68\\ 
    \bottomrule
    
  \end{tabular}
  \end{threeparttable}
  \begin{tablenotes}
      \item[1] * The time difference between when the UE sent a handover request and when the UE received the configuration response.
      \item[2] ** The time difference between when the UE sent a handover request and when the UE lost connectivity and did not receive the configuration response 
    \end{tablenotes}
    \vspace{-0.4cm}
\end{table*}

\vspace{1.5pt}\noindent\textbf{Handover scenario:} {Without loss of generality, we deploy three satellites} {with one logical cell each} to simulate a {generalized and} continuous {inter-satellite} handover scenario, as shown in Fig. \ref{fig:movingsat}. To better study the relationship between the density and handover performance, we randomly deploy 10,000 to 70,000 UEs and assign them to fixed square groups. We assume all UEs are connected to $SAT_1$ at the start of the simulation, and we report the performance of UE handover from $SAT_1$ to $SAT_2$. Our simulation also incorporates optimizations expected in real deployments, such as prioritizing messages, including messages from other satellites, responses from the core, and attachment requests from UE. Furthermore, we assume that a UE will require a handover when it is closer to another satellite, which is a location-based handover trigger~\cite{3gpp33821}. We ran the simulation five times with different seeds (10, 20, 30, 40, 50) to generate multiple UE deployment configurations. 

\vspace{1.5pt}\noindent\textbf{Baseline HO vs. GHO  protocol:}
In the baseline handover protocol (denoted HO) \cite{peltonen2021comprehensive, 3gpp.23.502}, a UE individually sends handover requests to the source satellite. Upon receiving the request, this satellite asks the target satellite for configuration data and then delivers it back to the UE. In our group handover protocol (denoted GHO), we allow the source satellite to send UE requests in batches to reduce message overhead. 

We ran simulations with both handover protocols, and Table~\ref{table:setting} shows the parameters we used for each protocol. For GHO, the ground space is partitioned into square areas, each with a fixed width. Within each square, UEs are organized into groups. Both HO and GHO are configured with a re-transmission timeout so that a UE will re-transmit the handover request after waiting a pre-defined time.

\subsection{Experimental Results}

\begin{figure}[t]
\centering
\includegraphics[width=0.75\linewidth]{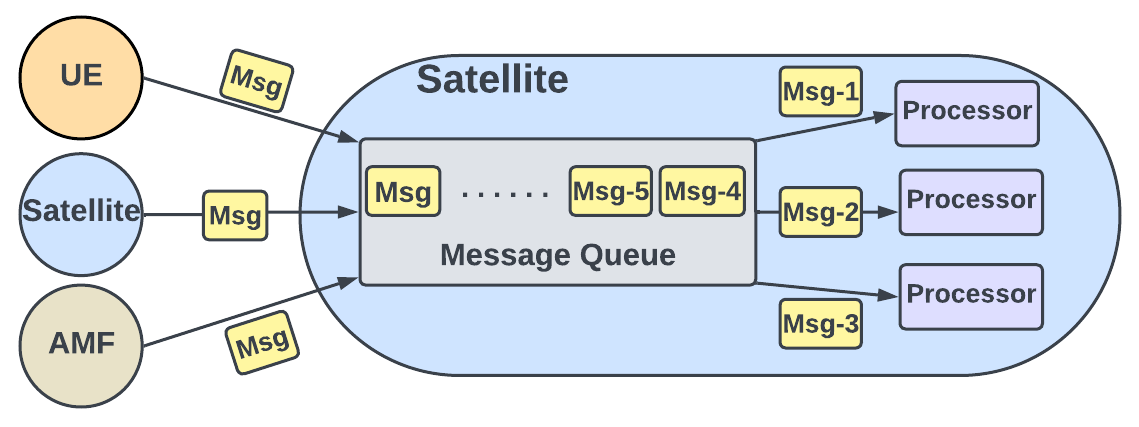}
\caption{Message processing model}
\vspace{-0.4cm}
\label{fig:queue}
\end{figure}

\begin{figure}[t]
\centering
\includegraphics[width=0.75\linewidth]{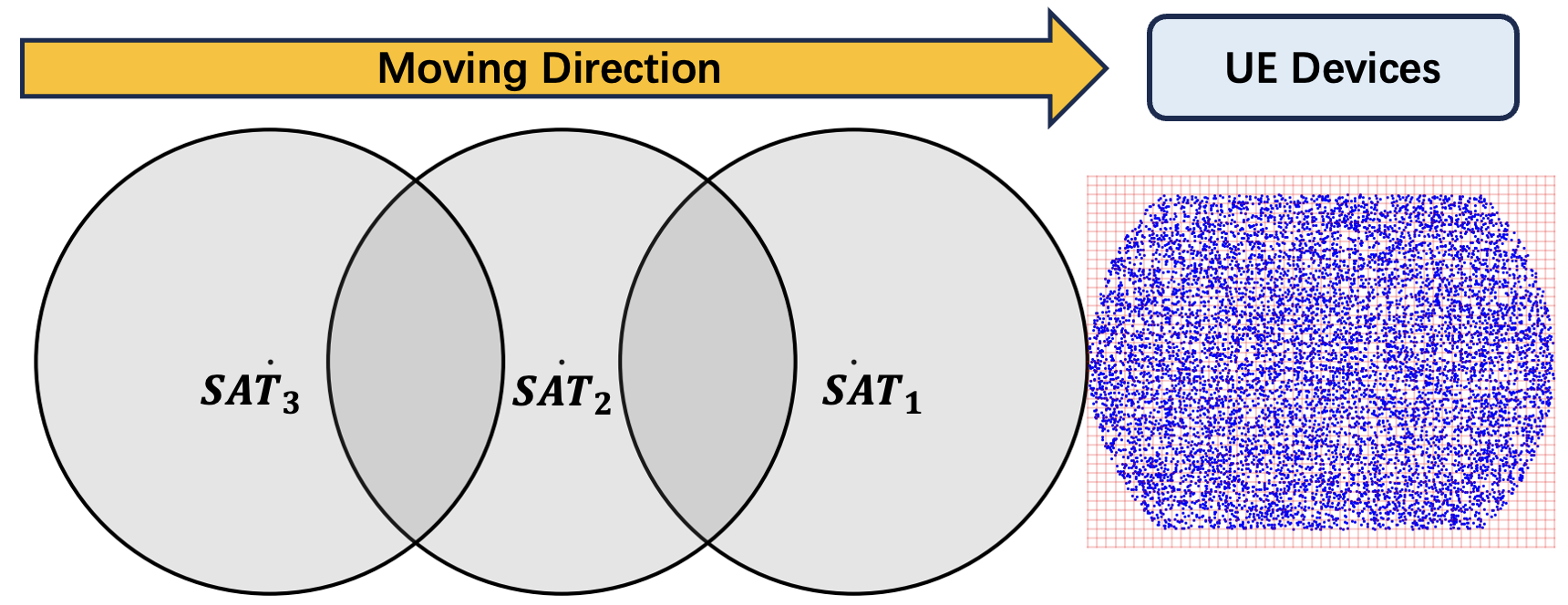}
\caption{Scenario of our experiment}
\label{fig:movingsat}
\end{figure}
Table \ref{table:res} shows the results of our experiment for $SAT_1$ in responding to handover requests. The success rate indicates the fraction of UEs that successfully complete the handover procedure. We also include the total messages the satellite received and the total UE handover/retransmit request messages. The drop rate indicates the number of dropped messages with respect to the total messages. UE waiting time is the difference between when UE sends a handover request or broadcasts the share and when the UE receives the configuration from the source satellite. Next, we discuss our main findings.

\vspace{1.5pt}\noindent\textbf{GHO achieves a higher successful handover rate:} In HO, the satellite is capable of processing the request when the number of UEs are 10,000 and 20,000. However, at 30,000 UEs, the UE waiting time increases, indicating that the UE started to re-transmit handover requests. We see the total received messages are increasing non-linearly, and the satellite drops messages, indicating a signaling storm overwhelms the satellite.
Consequently, from 30,000 UEs to 70,000 UEs, the handover success rate decreases, and the UE waiting time increases. For GHO, we see that the total messages are smaller than in HO, which reduces the burden on the satellite, and the success rate is much better than HO. However, at 60,000 UEs, GHO also experiences delay because $SAT_2$ needs to take over a massive number of UEs and delays the response to the messages from $SAT_1$, which is inevitable in a handover protocol. This is because a UE must individually access the satellite and achieve mutual trust using the new session key. We also observe an overall higher waiting time in GHO when the satellite is not overwhelmed.

\begin{figure}[t]
\includegraphics[width=0.92\linewidth]{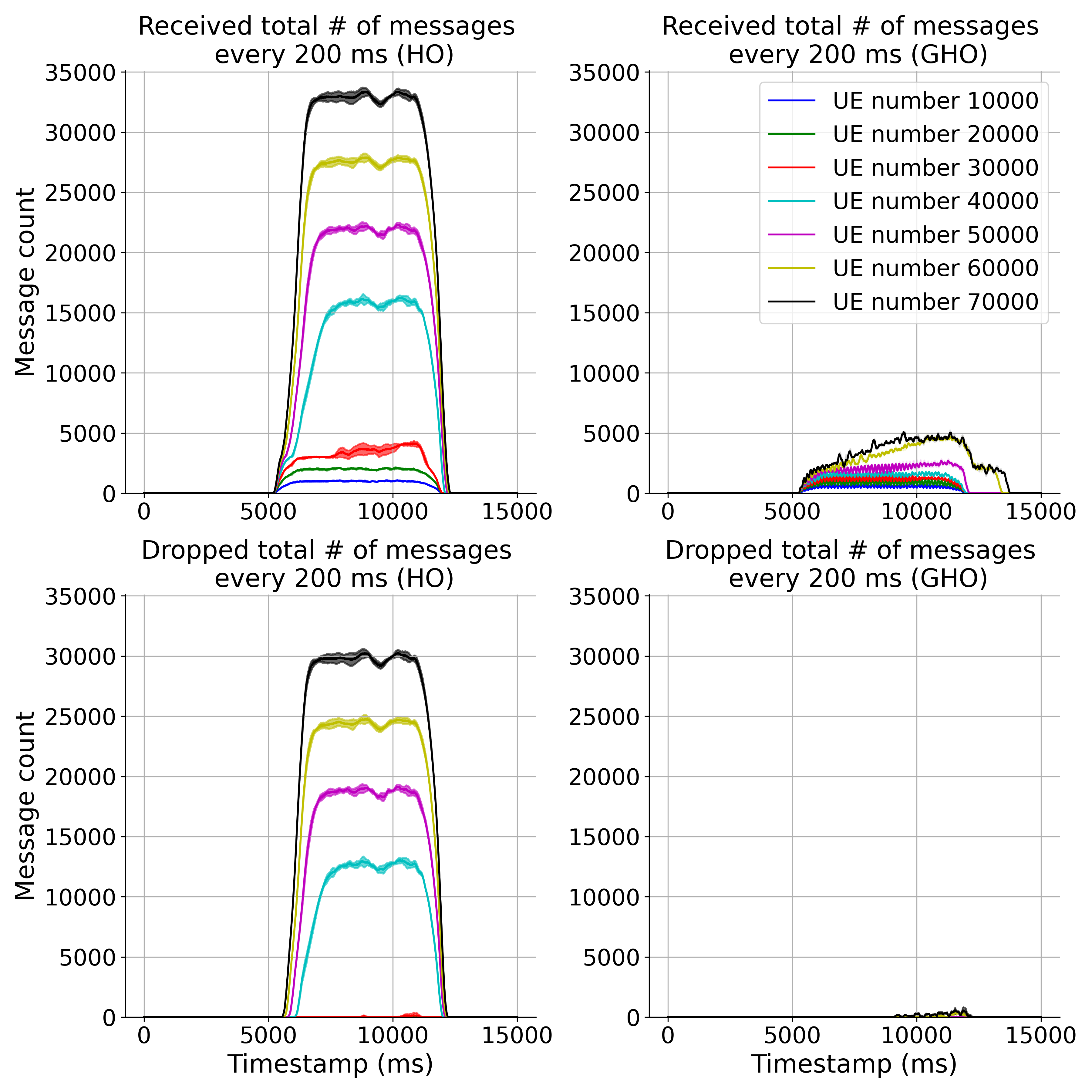}
\caption{Message count every 200 ms}
\label{fig:dynamic}
\end{figure}

\vspace{1.5pt}\noindent\textbf{GHO reduces satellites' message processing burden:} To analyze this dynamic process, we also collected the received and dropped messages every 200 ms through time. Fig. \ref{fig:dynamic} shows the result from $SAT_1$. Overall, due to the circular shape of a satellite's coverage, the total number of received messages and drop rate gradually increase and decrease. $SAT_1$ starts to handover UEs at 5,300 ms, and $SAT_2$ starts to handover UEs at 9,500 ms. We can see that when $SAT_1$ is overloaded and drops messages, the total number of messages dramatically increases because UE re-transmission messages amplify the signaling storm. Compared to HO, our GHO protocol significantly reduces the number of received messages. Even when the satellite is overloaded, as the number of $GAs$ is much smaller than the total UEs, the re-transmission messages do not cause a significant amplification of messages in  GHO. 

%However, we notice that when the $SAT_1$ is overloaded in GHO, the satellite still receives messages from AMF between 12000 ms and 14000 ms because the second satellite is overloaded and needs to prepare the broadcasting signature for group handover, which delays the UE attachment. 

%In HO, when UE number is 30000, the received UE requests increase at around 4000 ms but the satellite is not dropping the request and then increase at around 6000 ms with some drop rate. The reason for the first increase is natural because the satellite queued the tasks and some UEs re-transmit the message. The second increase is because the second satellite starts to handover to the third satellite while processing massive random access and it causes a late response to the first satellite.   

% Demonstate signaling storm

% Compare 

% We pick the following as the comparison KPI.
% \begin{enumerate}
%     \item Success handover Ratio.
%     \item The number of queued messages:
%         \begin{enumerate}
%             \item Messages from Satellite
%             \item Messages from AMF
%             \item Random access request messages from UEs
%             \item Handover request messages from UEs
%         \end{enumerate}
%     \item The number of dropped messages
    
%     \item Responding time delay: The time difference between request sending time and configuration receiving time.

%     \item Control Signal number.
% \end{enumerate}

  %\input{./content/7-discussion}
  \section{Conclusions}
\label{sec:conclusions}

We examined the signaling storm problem in 5G NTN handover and proposed a secure and efficient Xn-based GHO protocol to mitigate this issue. We implemented a discrete-event simulator to demonstrate that our GHO protocol helps reduce message overhead and handover additional UEs. However, we find that the overall handover process is still afflicted by a large number of access requests. {Formal security verification and evaluation with multi-cells will be performed in the future.} {We will also study GHO for NTNs with an Earth-fixed cell, where massive UEs in a cell will require handover simultaneously. Moreover, we will investigate signaling storm due to conditional handover protocol in 5G NTNs.}  
\section*{Acknowledgement}

We acknowledge the support of the NRC-Waterloo Collaboration Center (project reference number 090755). 
%\clearpage
\bibliographystyle{IEEEtran}
\bibliography{references}

\end{document}